\documentclass[sigconf]{acmart}

\usepackage{graphicx}
\usepackage{xspace}
\newcommand{\toolname}{{DetectBERT}\xspace}
\usepackage[normalem]{ulem}
\usepackage{xcolor}

\usepackage{booktabs}
\usepackage{siunitx}
\usepackage{hyperref}
\usepackage{amsmath} 
\usepackage{tcolorbox}
\usepackage{url}
\usepackage[normalem]{ulem}

\AtBeginDocument{%
  }

\copyrightyear{2024}
\acmYear{2024}
\setcopyright{rightsretained}
\acmConference[ESEM '24]{Proceedings of the 18th ACM / IEEE International Symposium on Empirical Software Engineering and Measurement}{October 24--25, 2024}{Barcelona, Spain}
\acmBooktitle{Proceedings of the 18th ACM / IEEE International Symposium on Empirical Software Engineering and Measurement (ESEM '24), October 24--25, 2024, Barcelona, Spain}\acmDOI{10.1145/3674805.3690745}
\acmISBN{979-8-4007-1047-6/24/10}

\begin{document}

\title{DetectBERT: Towards Full App-Level Representation Learning to Detect Android Malware}

\author{Tiezhu Sun}
\authornote{Corresponding Authors.}
\email{tiezhu.sun@uni.lu	}
\affiliation{%
  \institution{University of Luxembourg}
  \city{Kirchberg}
  \country{Luxembourg}
}

\author{Nadia Daoudi}
\email{nadia.daoudi@list.lu}
\affiliation{%
  \institution{Luxembourg Institute of Science and Technology}
  \city{Esch-sur-Alzette}
  \country{Luxembourg}
}

\author{Kisub Kim}
\authornotemark[1]
\email{falconlk00@gmail.com	}
\affiliation{%
  \institution{Independent Researcher}
  \city{Hongkong}
  \country{}
  }

\author{Kevin Allix}
\email{kallix@kallix.net}
\affiliation{%
  \institution{Independent Researcher}
  \city{Rennes}
  \country{France}
}

\author{Tegawendé F. Bissyandé}
\email{tegawende.bissyande@uni.lu}
\affiliation{%
  \institution{University of Luxembourg}
  \city{Kirchberg}
  \country{Luxembourg}
}

\author{Jacques Klein}
\email{jacques.klein@uni.lu	}
\affiliation{%
  \institution{University of Luxembourg}
  \city{Kirchberg}
  \country{Luxembourg}
}

\renewcommand{\shortauthors}{T. Sun et al.}

\begin{abstract}

Recent advancements in ML and DL have significantly improved Android malware detection, yet many methodologies still rely on basic static analysis, bytecode, or function call graphs that often fail to capture complex malicious behaviors. 
DexBERT, a pre-trained BERT-like model tailored for Android representation learning, enriches class-level representations by analyzing Smali code extracted from APKs. 
However, its functionality is constrained by its inability to process multiple Smali classes simultaneously.
This paper introduces DetectBERT, which integrates correlated Multiple Instance Learning (c-MIL) with DexBERT to handle the high dimensionality and variability of Android malware, enabling effective app-level detection. 
By treating class-level features as instances within MIL bags, DetectBERT aggregates these into a comprehensive app-level representation. 
Our evaluation demonstrates that DetectBERT not only surpasses existing state-of-the-art detection methods but also adapts to evolving malware threats.
Moreover, the versatility of the DetectBERT framework holds promising potential for broader applications in app-level analysis and other software engineering tasks, offering new avenues for research and development.

\end{abstract}

\maketitle

\section{Introduction}
The ubiquity of Android devices in today's digital ecosystem has made them a prime target for malicious actors.
As the sophistication of Android malware continues to evolve, so does the need for effective detection mechanisms. 
Traditional malware detection approaches, while beneficial, often struggle to keep pace with the rapid development of new and complex malware {apps}~\cite{qiu2020survey,chowdhury2023android}. 
This challenge is exacerbated by the dynamic and open nature of the Android platform, which allows for the frequent introduction of new applications and updates.

Recent advancements in machine learning~\cite{arp2014drebin,wu2019malscan} (ML) and deep learning~\cite{hsien2018r2,ding2020android,daoudi2021dexray} (DL) have offered promising directions for enhancing Android malware detection. 
These techniques, capable of learning from large datasets to identify subtle patterns and anomalies, have shown their potential across a broad range of software engineering applications.
However, they largely depend on static analysis~\cite{fereidooni2016anastasia,sandeep2019static,pan2020systematic}, extracting basic features or low-level bytecode from APK files~\cite{daoudi2021dexray,sun2021android}, and analyzing function call graphs~\cite{pektacs2020deep,yang2021android}. 
While these approaches showcase some effectiveness, they often fall short in capturing the nuanced and complex behaviors that characterize new emerging malware apps.

Recently, DexBERT~\cite{sun2023dexbert} has been proposed as a pre-trained BERT~\cite{devlin2018bert}-like model specifically designed for class-level Android representation learning.
By learning the Smali code disassembled from APKs, DexBERT has demonstrated its capability of significantly enhancing performance in class-level Android analysis tasks, such as malicious code localization. 
Nevertheless, the applications of DexBERT are limited by its input constraints, which makes it only able to analyze a single Smali class and cannot {support} an overall app-level understanding.

To address a similar issue in whole slide image classification and long document classification, TransMIL~\cite{shao2021transmil} and LaFiCMIL~\cite{sun2023laficmil} leverage correlated Multiple Instance Learning (c-MIL).
This enhances the ability to identify and learn valuable features from a vast array of cropped image patches and segmented document fragments, which are sourced from large medical images and lengthy documents, respectively.
Inspired by previous studies and recognizing the typical presence of multiple classes within an APK, we employ c-MIL principles for full app-level representation learning to detect Android malware, which led to the development of our proposed model, \toolname. 
Specifically, \toolname sets itself apart from TransMIL and LaFiCMIL by: 
\begin{itemize}
    \item Employing Smali classes directly as natural instances within the c-MIL framework, avoiding the need for preliminary data processing such as image cropping or document splitting.
    \item Omitting positional embeddings due to the non-sequential nature of Smali class interconnections, which are based on invocation links.
    \item Freezing DexBERT's weights during training to leverage the pre-trained model's original capabilities for generating class embeddings, thus conserving computational resources.
\end{itemize}

To implement c-MIL {for our scenario}, \toolname employs the Nyström Attention layer~\cite{xiong2021nystromformer}, a technique that efficiently learns the relationships among embedding vectors.
Specifically, upon generating class embeddings via DexBERT, we introduce a learnable category vector that is initialized to conform to a normal distribution and is dimensionally compatible with the class embeddings.
Subsequently, both the category vector and all class embeddings are processed through the Nyström Attention layer.
This mechanism helps \toolname learn the intricate correlations among the vectors, enabling a dynamic exchange of information between the category vector and each class embedding. 
Such an exchange is pivotal for highlighting features essential to malware detection. 
The process enriches the category vector, which is then passed through a fully connected layer, serving as the foundation for the ultimate malware detection decision.

We conduct an extensive evaluation on a large dataset of \num{158803} applications.
The experimental results demonstrate that \toolname not only surpasses three basic feature aggregation methods but also outperforms two state-of-the-art Android malware detection techniques.
Furthermore, we conduct a temporal consistency evaluation to assess how well \toolname performs over time, especially when confronted with newer malware samples absent in the training data. 
This assessment reaffirms the robustness of \toolname, demonstrating its sustained effectiveness against emerging threats and highlighting its relevance in dynamic real-world scenarios.
This research bridges the gap between sophisticated representation learning models and the practical demands of app-level malware analysis. 
Looking forward, \toolname showcases significant potential to expand its utility in software engineering, particularly by using Multiple Instance Learning (MIL) to efficiently process and analyze large-scale inputs. 
This capability is crucial for advancing software quality and security assessments in complex environments like mobile and IoT ecosystems.

The contributions of our study are as follows: 
\begin{itemize}
    \item We introduce \toolname, an efficient and effective approach utilizing MIL to scale DexBERT for full app-level representation learning in Android malware detection. 
    \item We perform a comprehensive evaluation, demonstrating \toolname's superiority over both basic feature aggregation methods and state-of-the-art malware detection techniques, including its robustness through a temporal consistency evaluation to verify performance against new, unseen malware samples.
    \item We highlight \toolname's potential to revolutionize software engineering, specifically through its use of MIL to manage large-scale data, offering significant improvements in software assessments.
    \item To facilitate replication, we have made the dataset and source code available at: \\
    \url{https://github.com/Trustworthy-Software/DetectBERT}
\end{itemize}

\section{Background}
To provide a foundation for understanding this study, this section delves into the background of three core concepts: Android malware detection, DexBERT, and multiple instance learning.

\subsection{Android Malware Detection}
Traditional Android malware detection methods have increasingly fallen short in addressing the complexity of malware varieties~\cite{qiu2020survey,chowdhury2023android}. The integration of machine learning~\cite{arp2014drebin,wu2019malscan} and deep learning techniques~\cite{hsien2018r2,ding2020android,daoudi2021dexray} into the field represents a significant advancement, promising enhanced detection capabilities.
Nonetheless, the application of these sophisticated approaches in Android malware detection encounters substantial challenges, as they largely rely on static analysis~\cite{fereidooni2016anastasia,sandeep2019static,pan2020systematic}, low-level bytecode~\cite{daoudi2021dexray,sun2021android}, and function call graphs~\cite{pektacs2020deep,yang2021android}, potentially overlooking the intricate behaviors inherent in modern malware.
\toolname is the fist endeavor to enable DexBERT~\cite{sun2023dexbert} to detect Android malware with a comprehensive app-level representation, from a novel perspective of multiple instance learning.

\subsection{DexBERT}

DexBERT~\cite{sun2023dexbert} is a pretrained model based on the BERT architecture~\cite{devlin2018bert}.
It is designed specifically for extracting class-level features from Android applications.
It processes disassembled Smali code from Dalvik bytecode as input and learns to produce corresponding representations, i.e., the embedding of a Smali class. Smali serves as the textual representation of Dalvik bytecode, similar to how assembly code represents compiled code in a readable form.
These class embeddings, or feature vectors, enable the execution of various class-level Android analysis tasks, including malicious code localization. 
DexBERT's ability to understand the intrinsic logic and behavior of applications represents a substantial step forward in the battle against malware.
However, its input capacity is limited to 512 BERT tokens, which is only enough to process one single Smali class. 
Considering that an APK typically contains thousands of Smali classes, there is a clear need for an effective class embedding aggregation method. 
Our proposed method, \toolname, is designed to leverage DexBERT's representation learning capabilities for app-level tasks, such as Android malware detection.

\begin{figure*}[ht]
\includegraphics[width=\textwidth]{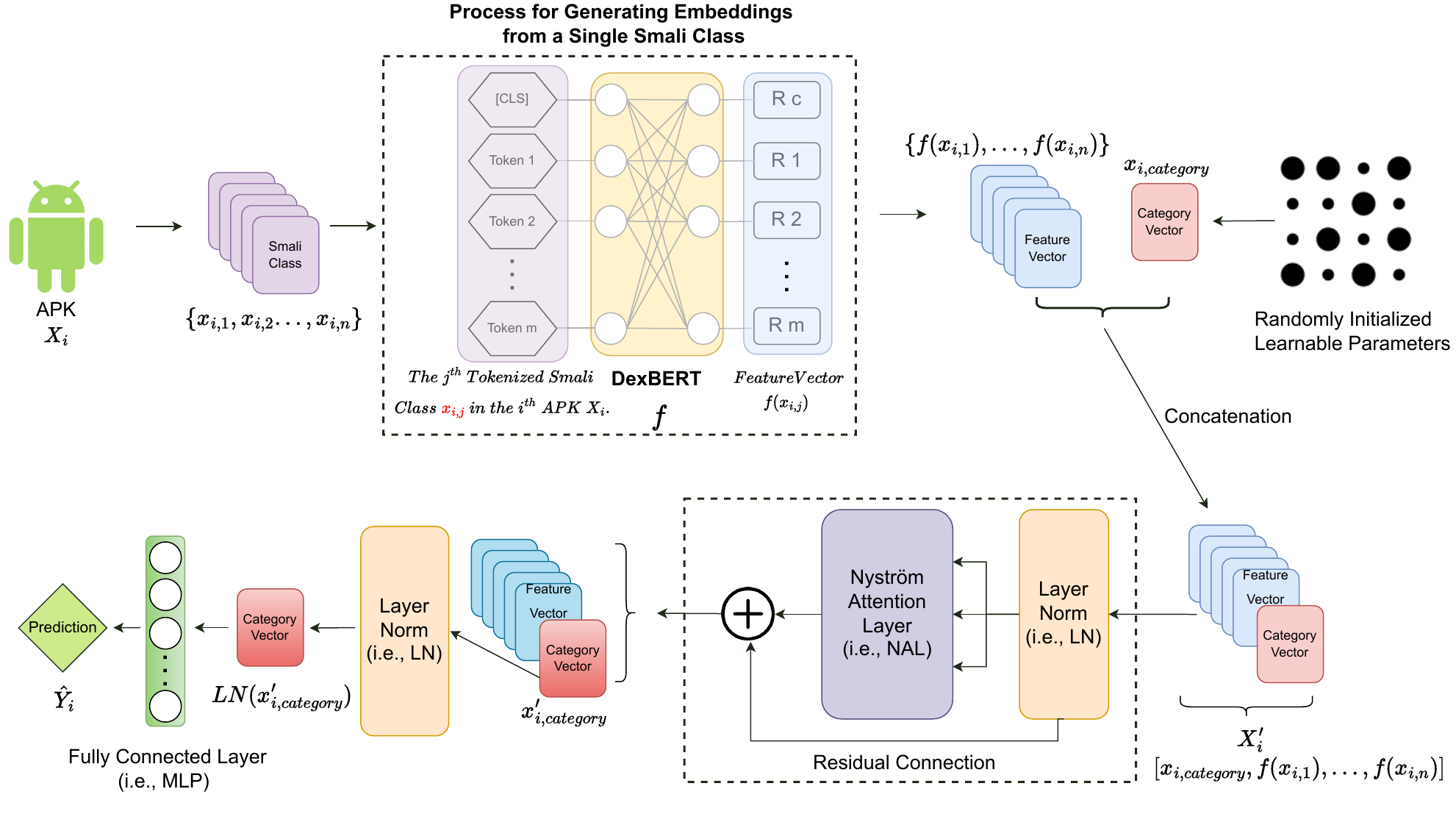}
\caption{Overview of \toolname Workflow. First, DexBERT produces Smali class embeddings as c-MIL instances. A category vector of the same size is then introduced as an additional instance. The Nyström Attention layer helps \toolname find correlations among instances, allowing the category vector to capture key information from class embeddings for malware detection. Lastly, this vector is processed in a fully connected layer to make the detection decision.}
\label{fig:overview}
\end{figure*}

\subsection{Multiple Instance Learning}

In this study, we adopt Multiple Instance Learning~\cite{ji2020diversified,song2019using,hebbar2021deep} (MIL) as our strategy for integrating class-level embeddings into an overall app-level representation.
MIL is characterized by its ability to handle a collection of instances grouped into a ``bag'', where the bag is assigned a single label, but individual instance labels are not provided. 
A notable challenge within MIL is the variable number of instances each bag may contain, requiring a MIL model that can adeptly manage bags of diverse sizes.
MIL's versatility has been demonstrated in various domains, especially in computer vision~\cite{shao2021transmil,zhang2022dtfd} and in natural language processing~\cite{ji2020diversified,sun2023laficmil}. 
These successful applications inspired us to explore MIL's potential in enhancing Android malware detection using DexBERT. 
Specifically, by conceptualizing each APK as a ``bag'' and its constituent Smali classes as ``instances'' within this framework, we are able to establish a fitting model for detecting malware in Android applications. 
However, standard MIL may not fully capture the complex interconnections between Smali classes. This is where correlated Multiple Instance Learning (c-MIL) comes in, which considers the dependencies and interactions among instances within a bag. c-MIL enables \toolname to assess collective class behavior, crucial for identifying Android malware, by recognizing patterns that emerge from the invocation relationships among classes.

\section{Approach}
In this section, we start with the key theoretical foundations of our method, then detail the design of \toolname.

\subsection{Theoretical Foundations}
\label{sec:theorem}

\begin{theorem}
\label{theorem:approximation}
Suppose $S : \chi \rightarrow \mathbb{R}$ is a continuous set function w.r.t Hausdorff distance~\cite{rote1991computing} $d_{H}(., .)$. $\forall \varepsilon > 0$, for any invertible map $P : \chi \rightarrow \mathbb{R}^{n}$, $\exists$ function $\sigma$ and $g$, such that for any set $X \in \chi$: 

\begin{equation}
\small
    |S(X) - g(P_{X \in \chi}\{\sigma(x): x \in X\})| < \varepsilon
\end{equation}
\end{theorem}

The proof of Theorem~\ref{theorem:approximation} can be found in~\cite{shao2021transmil}.
This theorem shows that a Hausdorff continuous {\bf set function} $S(X)$ can be arbitrarily approximated by a function in the form $g(P_{X \in \chi}\{\sigma(x) : x \in X\})$.
This concept applies to MIL, where the theorem's sets correspond to MIL's bags.
Consequently, the theorem provides a foundation for approximating bag-level predictions in MIL using instance-level features.
In the context of Android malware detection, this principle instructs us to achieve app-level predictions through the utilization of class-level embeddings produced by DexBERT.

\begin{theorem}
\label{theorem:entropy}
The instances in the bag are represented by random variables $\theta_{1}, \theta_{2}, ..., \theta_{n}$, the information entropy of the bag under the correlation assumption can be expressed as $H(\theta_{1}, \theta_{2}, ..., \theta_{n})$, and the information entropy of the bag under the i.i.d. (independent and identical distribution) assumption can be expressed as $\sum_{t=1}^{n} H(\theta_{t})$, then we have:
\begin{equation}
\small
\begin{split}
H(\theta_{1}, \theta_{2}, ..., \theta_{n}) &= \sum_{t=2}^{n} H(\theta_{t} | \theta_{1}, \theta_{2}, ..., \theta_{t-1}) + H(\theta_{1}) \\ & \leq \sum_{t=1}^{n} H(\theta_{t})
\end{split}
\end{equation}
\end{theorem}
The proof of Theorem~\ref{theorem:entropy} can be found in~\cite{shao2021transmil}.
This theorem demonstrates that a bag under the correlation assumption has lower information entropy than that under the i.i.d. assumption (i.e., in standard MIL).
The lower information entropy in c-MIL suggests reduced uncertainty and the potential to provide more valuable information for bag classification tasks than the standard MIL.
In the context of Android malware detection, we claim that the Smali classes from the same APK are correlated in some way (e.g., invocation relationships between classes). This implies that the presence or absence of a malicious class in a bag can be influenced by the other classes contained within the APK.
Therefore, c-MIL seems a perfect choice to leverage the class embeddings of DexBERT for the app-level task of Android malware detection.

Given an app $X_{i}$ composed of Smali classes $\{x_{i,1}, x_{i,2}, ..., x_{i,n}\}$, for $i=1, ..., N$, that exhibit correlation among each other.
The app-level label is $Y_{i}$, yet the class-level labels $\{y_{i,1}, y_{i,2}, ..., y_{i,n}\}$ are not accessible.
Then, detecting Android malware in the context of c-MIL can be defined as:
\begin{equation}
\small
Y_{i} = \left\{\begin{array}{l}
0, \indent if \sum y_{i,j}=0 \indent y_{i,j}\in \{0, 1\}, j=1,...,n\\ 
1, \indent otherwise
\end{array}\right.
\end{equation}

\begin{equation}
\small
\label{eq:score}
    \hat Y_{i} = S(X_{i}),
\end{equation}
where $S$ is a scoring function and $\hat Y$ is the predicted score indicating the likelihood of the analyzed application being malware.
$N$ is the total number of applications, and $n$ is the number of Smali classes in the $i$th application. 
The number $n$ generally varies for different applications.

\subsection{DetectBERT}
\label{sec:detectbert}

According to Theorem~\ref{theorem:approximation}, we leverage a series of {\bf sub-functions to approximate} the c-MIL score function $S$ defined in Equation~\ref{eq:score}. 
Given a set of APKs $\{X_{1}, ..., X_{N}\}$, where each APK $X_{i}$ contains multiple Smali classes $\{x_{i,1}, ..., x_{i,n}\}$ and an APK label $Y_{i}$, a corresponding category vector $x_{i, category}$ is randomly initialized.
We represent DexBERT~\cite{sun2023dexbert} as a feature extracting function $f$, Multi-layer Perceptron~\cite{rumelhart1986learning} as $MLP$, Nyström Attention Layer~\cite{xiong2021nystromformer} as $NAL$ and Layer Normalization~\cite{ba2016layer} as $LN$. The c-MIL score function $S$ can be approximated as follows:
\begin{equation}
\label{eq:vecs}
    X'_{i} = [x_{i, category}, f(x_{i,1}), ..., f(x_{i,n})]
\end{equation}

\begin{equation}
    x'_{i, category} = (NAL(LN(X'_{i})) + X'_{i})^{(0)}
\end{equation}

\begin{equation}
    \hat Y_i = MLP(LN(x'_{i, category}))
\end{equation}

The selection of the Nyström Attention layer in \toolname's architecture is driven by its capability to efficiently handle large sequences of data.
First, as an attention mechanism, Nyström Attention naturally facilitates the learning of correlations among class embeddings, treating them akin to tokens in natural language processing. This enables \toolname to dynamically assess and prioritize the relevance of each Smali class' features based on their contextual relationship, crucial for accurate malware detection. 
Secondly, when confronted with a vast number of tokens—typical in complex Android applications—the Nyström Attention layer proves significantly more efficient than traditional attention mechanisms. Its design reduces computational complexity from quadratic to linear, making it especially suitable for large-scale datasets where efficiency is paramount. These characteristics make Nyström Attention an ideal choice for \toolname. 

\toolname distinguishes itself from TransMIL and LaFiCMIL in three key aspects. 
Firstly, unlike the image cropping in TransMIL or document splitting in LaFiCMIL, \toolname directly utilizes Smali classes as natural instances for c-MIL without requiring any preprocessing steps. 
Secondly, our model does not employ positional embeddings, as outlined in Equation~\ref{eq:vecs}. This decision is based on the observation that Smali classes are interconnected through invocation relationships rather than a sequential order. 
Lastly, to enhance efficiency, the weight parameters of DexBERT are frozen during training. \toolname leverages the DexBERT’s original capacity to generate meaningful class embeddings without the need for further fine-tuning, thus avoiding additional computational costs.

The overall process of \toolname is outlined in Figure~\ref{fig:overview} and can be detailed as follows: given an APK $X_i$ containing $n$ Smali classes $\{x_{i,1}, ..., x_{i,n}\}$, we use the pretrained DexBERT model $f$ to generate the corresponding embedding vectors $\{f(x_{i,1}), ..., f(x_{i,n})\}$. 
Then, we initialize a {\bf learnable} category vector $x_{i, category}$ that conforms to a normal distribution and has the same shape as each class embedding.
By considering the category vector as an additional embedding vector, we concatenate it with the class embeddings as $[x_{i, category}, f(x_{i,1}), ..., f(x_{i,n})]$, and learn the correlation between each embedding vector using the Nyström Attention layer.
Layer normalization and residual connection~\cite{he2016deep} are applied to standardize the input and output data of the Nyström Attention layer and facilitate gradient backpropagation, enhancing model stability and performance.
With the help of the attention mechanism, the category vector exchanges information with each class embedding and extracts necessary features for malware detection.
Finally, the category vector is fed into a fully connected layer (i.e., MLP) to finalize the detection task.

\section{Study Design}
This section outlines the design of our study, detailing the research questions, dataset, empirical setup, and evaluation metrics used to assess the performance of \toolname. 

\subsection{Research Questions}
\label{sec:RQs}
We mainly investigate three research questions, each aimed at exploring different aspects of \toolname's performance:

\begin{itemize}
    \item {\bf RQ1}: How does \toolname perform compared to basic feature aggregation methods in detecting Android malware?
    \item {\bf RQ2}: How does \toolname perform compared to state-of-the-art Android malware detection models?
    \item {\bf RQ3}: How does \toolname maintain its detection effectiveness over time in the face of evolving Android malware?
\end{itemize}

These questions are designed to comprehensively evaluate the effectiveness, competitiveness, and robustness of \toolname. 

\subsection{Dataset}

For our evaluation, we utilize the large-scale benchmark dataset from DexRay~\cite{daoudi2021dexray}, which contains a total of \num{158803} apps. 
This dataset facilitates a direct comparison of \toolname against basic aggregation methods and the state-of-the-art Android malware detection approaches.
The malware and benign apps in the DexRay dataset are selectively collected from the popular and continually growing Android repository named AndroZoo~\cite{allix2016androzoo}. 
Specifically, the DexRay dataset contains \num{96994} benign and \num{61809} malware apps.
Benign apps are defined as the apps that have not been detected by any antivirus from VirusTotal~\footnote{\url{https://www.virustotal.com/}}, while the malware collection contains the apps that have been detected by at least two antivirus engines.

\subsection{Empirical Setup}
\label{sec:emp_setup}

In our empirical setup, we employ the DexBERT model to generate a feature vector for each Smali class. Considering the extensive size of the dataset, we optimize our process by storing these generated feature vectors on disk, thereby significantly reducing computational costs by preventing the need to regenerate feature vectors during each training phase. Consequently, the \toolname model requires only 2 GB of GPU memory for training and achieves an average inference time of just 0.005 seconds per app, underscoring its computational efficiency.

To enhance the model's capacity without altering the underlying DexBERT architecture, we freeze DexBERT's parameters and incorporate dual Nyström Attention layers. 
This approach allows us to expand the model's capabilities while maintaining a stable structure.
For the optimization process, we employ the Lookahead optimizer~\cite{zhang2019lookahead} to fine-tune the model. The training is conducted for 20 epochs with a learning rate set to \texttt{1e-4}.

\subsection{Evaluation Metrics}
\label{sec:metrics}
For comparison with basic aggregation methods in Section~\ref{sec:basic-comparison} and comparison with state-of-the-art approaches in Section~\ref{sec:sota-comparison}, we follow the baseline DexRay~\cite{daoudi2021dexray} by shuffling the dataset and split it into 80\% for training, 10\% for validation, and 10\% for test. 
This process is repeated 10 times to ensure robustness, with the average results reported.
In Section~\ref{sec:temp}, to assess temporal consistency, we temporally partition our dataset, using apps from 2019 for training (90\%) and those from 2020 for testing (10\%), based on the date information in the app's DEX file.
Results are reported using the four metrics: accuracy (\%), precision (\%), recall (\%), and F1 Score (\%), with precision maintained to two decimal places for consistency with baseline results.
\section{Experimental Results}
In this section, we evaluate \toolname's performance by addressing the three research questions outlined in Section~\ref{sec:RQs}, aiming to provide insights into distinct aspects of its effectiveness.

\subsection{RQ1: How does \toolname perform compared to basic feature aggregation methods in detecting Android malware?}
\label{sec:basic-comparison}

In this subsection, we evaluate \toolname's performance in detecting Android malware by comparing it against three fundamental techniques for aggregating DexBERT embeddings: Random Selection, Element-wise Addition, and Element-wise Average.
Each technique compresses information from multiple class embeddings into a single representative vector, which is then used as input to a fully connected layer for final prediction.

\begin{itemize}
    \item \textbf{Random Selection}: It randomly selects a single class embedding to represent the entire APK, testing the efficacy of individual class features in malware detection.
    \item \textbf{Element-wise Addition}: It aggregates class embeddings by adding corresponding elements together, aiming to capture the cumulative effect of features across all classes.
    \item \textbf{Element-wise Average}: Similar to addition, it calculates the average of corresponding elements, offering a normalized representation of class features.
\end{itemize}

\begin{table}[h] 
\centering
\caption{Performance comparison with basic feature aggregation approaches. } 
\label{tab:aggregation}
\scalebox{0.9}{%
\begin{tabular}{lccccc}
\toprule
Model &Accuracy &Precision &Recall &F1 Score \\
\midrule
Random Selection &0.82 &0.82 &0.82 &0.82 \\
Element-wise Addition &0.86 &0.87 &0.84 &0.86 \\
Element-wise Average &0.92 &0.92 &0.92 &0.92 \\
\midrule
\toolname & \textbf{0.97} &\textbf{0.98} &\textbf{0.95} & \textbf{0.97} \\
\bottomrule
\end{tabular}
}
\end{table}

Table~\ref{tab:aggregation} shows that while the Element-wise Average method achieves an F1 score of 0.92, indicating the effectiveness of DexBERT embeddings, \toolname significantly outperforms this and other basic methods. 
Notably, \toolname attains superior precision and F1 scores of 0.98 and 0.97, respectively, improving performance by 6 and 5 percentage points over the Element-wise Average method.

This substantial improvement is attributed to \toolname’s sophisticated architecture, which more effectively captures and utilizes the intricate relationships among class embeddings. Unlike basic aggregation methods that compress embeddings into a singular vector, \toolname employs Nyström Attention layers. These layers dynamically adjust and integrate information from different classes based on their contextual relevance, facilitating a more nuanced aggregation. This advanced capability proves crucial in detecting Android malware, where subtle interactions among app features often signify malicious activity.

\begin{tcolorbox}[colback=gray!5!white,colframe=gray!75!black,title=RQ1 Answer:]
\toolname significantly outperforms basic aggregation methods in detecting Android malware, emphasizing the importance of advanced embedding processing techniques for handling complex Android APKs, thus enhancing the reliability and accuracy of malware detection systems. 
\end{tcolorbox}

\subsection{RQ2: How does \toolname perform compared to state-of-the-art Android malware detection models?}
\label{sec:sota-comparison}

In this subsection, we evaluate the competitive performance of \toolname against two established state-of-the-art models in the field of Android malware detection: Drebin~\cite{arp2014drebin} and DexRay~\cite{daoudi2021dexray}. Drebin is known for its application of machine learning techniques to hand-crafted features, while DexRay utilizes a deep learning framework to analyze low-level bytecode images.

\begin{table}[h] 
\centering
\caption{Performance comparison with existing state-of-the-art approaches.} 
\label{tab:android}

\begin{tabular}{lccccc}
\toprule
Model &Accuracy &Precision &Recall &F1 Score \\
\midrule
Drebin &0.97 &0.97 &0.94 &0.96 \\
DexRay &0.97 &0.97 &0.95 &0.96 \\
\midrule
\toolname & \textbf{0.97} &\textbf{0.98} &\textbf{0.95} & \textbf{0.97} \\
\bottomrule
\end{tabular}

\end{table}

According to the findings in Table~\ref{tab:android}, \toolname not only matches but also exceeds the precision and F1 scores of the well-established Drebin and DexRay models on the challenging DexRay dataset. It elevates the precision score to 0.98, improving upon the already high score of 0.97 achieved by both Drebin and DexRay while maintaining high recall. 
These slight yet significant improvements in precision and F1 scores are critical as they highlight \toolname’s exceptional ability to pinpoint malware accurately without missing genuine threats. 
This performance enhancement is primarily due to \toolname's effective use of sophisticated c-MIL mechanisms. 
These mechanisms dynamically adjust based on the relationships among class embeddings, enabling \toolname to capture subtle nuances and connections within the Smali classes. 
This capability distinguishes it from other models that rely on conventional feature analysis techniques and may overlook such intricate relationships.
This nuanced detection capability is key to \toolname's success, offering deeper insights into the complex behaviors typical of modern malware.

\begin{tcolorbox}[colback=gray!5!white,colframe=gray!75!black,title=RQ2 Answer:]
\toolname slightly surpasses the performance of state-of-the-art models Drebin and DexRay. 
This superior performance still demonstrates the effectiveness of its architecture in leveraging MIL to enhance detection accuracy in complex malware datasets.
\end{tcolorbox}

\subsection{RQ3: How does \toolname maintain its detection effectiveness over time in the face of evolving Android malware?}
\label{sec:temp}

Model aging~\cite{xu2019droidevolver,zhang2020enhancing} is a significant challenge in machine learning, where a model's performance tends to degrade when applied to new, previously unseen samples.
In this section, we assess the robustness of \toolname against model aging by evaluating its temporal consistency, a measure of how well a model trained on historical data performs against recent threats.
The methodology for this temporal consistency evaluation is detailed in Section~\ref{sec:metrics}, where the dataset is divided into training sets from 2019 and testing sets from 2020 to accurately emulate real-world applications against evolving malware. The performance of \toolname and other state-of-the-art models is compared using these temporally distinct datasets.

\begin{table}[h]
\centering
\caption{Temporal consistency performance comparison with state-of-the-art approaches.} 
\label{tab:temp}

\begin{tabular}{lccccc}
\toprule
Model &Accuracy &Precision &Recall &F1 Score \\
\midrule
Drebin &0.96 &0.95 &0.98 &0.97 \\
DexRay &0.97 &0.97 &0.98 &0.98 \\
\midrule
\toolname & \textbf{0.99} &\textbf{0.99} &\textbf{0.99} & \textbf{0.99} \\
\bottomrule
\end{tabular}

\end{table}

Table~\ref{tab:temp} showcases the evaluation results, indicating that all models tested exhibit commendable adaptability to new malware samples. 
Notably, \toolname excels, achieving superior accuracy, precision, recall, and F1 scores of 0.99 across all metrics. 
This performance not only surpasses the other state-of-the-art models but also demonstrates significant improvements over the scores reported in Table~\ref{tab:android}, where the earlier data sample composition varied.
This discrepancy highlights the exceptional capability of \toolname to generalize from past data to future scenarios without significant performance degradation. 
The increased proportion of training data from 2019 in this setup (90\% compared to the previous 80\%) likely played a significant role in enhancing the robustness of \toolname. 
This broader learning base enabled it to more effectively identify malware characteristics that persist or evolve over time, contributing to its heightened effectiveness.

\toolname’s standout performance in this temporal consistency evaluation confirms its effectiveness in handling new malware threats, a key advantage for maintaining relevance in the rapidly evolving landscape of Android security. 
This high level of adaptability stems from \toolname’s sophisticated representation learning capabilities, which extract and leverage the high-level semantics from Smali code in Dex files. 
Unlike simpler models that might rely on surface-level features, \toolname deeply understands the underlying behaviors and patterns encoded in the application’s code, allowing it to detect even subtly disguised malware.
This adaptability is critical for practical deployment, where the ability to perform consistently over time is paramount.

\begin{tcolorbox}[colback=gray!5!white,colframe=gray!75!black,title=RQ3 Answer:]
\toolname exhibits exceptional temporal consistency in malware detection, effectively mitigating model aging issues and demonstrating robust generalization capabilities to new and evolving threats. 
These qualities position our model as an invaluable asset in the ongoing battle against Android malware.
\end{tcolorbox}
\section{Conclusion}
In this paper, we introduce \toolname, a novel framework that leverages the capabilities of DexBERT for app-level analysis, specifically tailored for the complex task of Android malware detection. 
By employing a correlated Multiple Instance Learning (c-MIL) strategy, \toolname not only surpasses traditional feature aggregation methods but also outperforms current state-of-the-art malware detection models. 
The performance improvement suggests that \toolname could set a new standard for Android security analysis, representation learning, and other software engineering tasks.

\textbf{Future Work:} We aim to further validate \toolname by comparing it with additional malware detection models and enhancing its ability to analyze a wider range of malware behaviors. 
This includes integrating continuous learning mechanisms and incorporating more comprehensive contextual data from APKs. 
Furthermore, we are excited about the prospects of adapting our c-MIL-based framework for other software engineering tasks that handle large-scale data inputs, potentially broadening its applicability and impact in the field.

\section*{Acknowledgment}
This research was funded in whole, or in part, by the Luxembourg National Research Fund (FNR), grant references 16344458 (REPROCESS), 18154263 (UNLOCK), and 17046335 (AFR PhD grant).


\bibliographystyle{ACM-Reference-Format}
\bibliography{sections/references}



\end{document}